\documentclass[12pt]{article}
\usepackage{mathtools,latexsym,color,graphicx,amsmath,amssymb,amsthm,url}
\usepackage{amsfonts}
\usepackage{euscript}

\def\dist{\varrho}
\def\polylog{\mathop{\mathrm{polylog}}}

\def\bn{{\bf bn}}

\def\marrow{\marginpar[\hfill$\longrightarrow$]{$\longleftarrow$}}
\def\micha#1{\textsc{(Micha says: }\marrow\textsf{#1})}

\newcommand{\old}[1]{{{}}}

\newtheorem{theorem}{Theorem}[section]

\makeatletter
\long\def\@makecaption#1#2{
   \vskip 10pt
   \setbox\@tempboxa\hbox{{\footnotesize {\bf #1.} #2}}
   \ifdim \wd\@tempboxa >\hsize         
       {\footnotesize {\bf #1.} #2\par}
     \else                              
       \hbox to\hsize{\hfil\box\@tempboxa\hfil}
   \fi}
\makeatother

\title{Efficient Algorithms for the Bottleneck Path Problem in Geometric Graphs\thanks{%
  Work on this paper has been partially supported by ISF Grant 495/23. 
  }}

\author{Matthew J. Katz\thanks{%
   The Stein Faculty of Computer and Information Science, Ben-Gurion University of the Negev, Beer Sheva, Israel; 
   {\sf matya@bgu.ac.il}; 
   {\sf https://orcid.org/0000-0002-0672-729X.}}
 \and
 Rachel Saban\thanks{%
   The Stein Faculty of Computer and Information Science, Ben-Gurion University of the Negev, Beer Sheva, Israel; 
   {\sf rachelfr@post.bgu.ac.il}}
 \and
 Micha Sharir\thanks{%
   School of Computer Science, Tel Aviv University, Tel Aviv Israel;
   {\sf michas@tauex.tau.ac.il};
   {\sf https://orcid.org/0000-0002-2541-3763.}}}

\begin{document}
\maketitle

\begin{abstract}
We present efficient algorithms for the bottleneck path problem in two geometric settings that arise naturally in applications: directional-antenna graphs in the plane with antenna angles bounded from below by a constant, and visibility graphs whose vertices lie on or above a 1.5-dimensional terrain, both with Euclidean distances as edge weights.
We provide near-linear algorithms for the corresponding decision problems, namely, 
determining whether the subgraph obtained by retaining all edges with weight at most some threshold $\bn$ contains a path from $s$ to $t$. 
We then use the decision procedures to obtain algorithms for the bottleneck path problem that run in $O^*(n^{8/7})$
randomized expected time, where $n$ is the input size and
the $O^*(\cdot)$ notation hides subpolynomial factors.

Within the same performance bounds, we can also solve the bounded-hop version, in which we only consider $s$-$t$ paths
with at most $k$ edges, for a given integer $k < n$.

\end{abstract}

\newpage
\section{Introduction}
\label{sec:intro}

Let $G=(V,E)$ be a weighted (and possibly directed) graph, and let $w$ be a positive weight function on the edge set $E$. 
The \emph{bottleneck} of a path $\pi$ in $G$, denoted $\bn(\pi)$, is the weight of the heaviest edge of $\pi$. In the 
\emph{bottleneck path} problem, we are given two vertices $s,t \in V$, and the goal is to find a path from $s$ to $t$ with 
minimum bottleneck (assuming there exists a path from $s$ to $t$ in $G$). We denote the bottleneck of such an optimal path
by $\bn(s,t)$, that is, $\bn(s,t) = \min \{\bn(\pi) \,|\, \pi \mbox{ a path in } $G$ \mbox{ from } $s$ \mbox{ to } $t$\}$.

Consider the associated \emph{decision problem}, which receives as input a threshold $\bn$, and aims to determine whether $G$ contains a path $\pi$ from 
$s$ to $t$ with $\bn(\pi) \le \bn$, that is, whether
$G_{\le \bn} = \{(u,v)\in G \mid w(u,v) \le \bn\}$ has a path from $s$ to $t$. In the problems that we study,
we are able to solve that problem in near-linear time, but we do not know how to achieve such a running time for the 
bottleneck path problem itself; see below how we tackle such situations. 

We also consider the \emph{bounded-hop} version of the problem, in which we are also given an integer parameter
$1 \le k < n$, and wish to determine, in the decision version, whether $G_{\le \bn}$ contains a path from $s$ to $t$ with at most $k$ edges. In the
optimization version, i.e., the bottleneck path problem, we seek the minimum value of $\bn$ for which $G_{\le \bn}$ 
contains a path from $s$ to $t$ with at most $k$ edges. This is also known as the \emph{reverse shortest path} (RSP)
problem for $G$.

For example, consider the case where $G$ is a Euclidean graph (some suitably defined subgraph of the complete graph) 
on a set $P$ of $n$ points in the plane. In this case we can sometimes solve the decision problem in near-linear time 
(it is not always an easy task, and depends on how $G$ is specified), but even in such cases, when the constraints
defining $G$ are nontrivial, we do not know how to solve 
the bottleneck path problem (the optimization version of the problem) in near-linear time, as it is
generally based on running BFS on the threshold graph, and we do not know how to simulate the execution of BFS in parallel,
in the standard style of parametric search.\footnote{%
  Running BFS in small parallel depth is a general notorious open problem.}
A simple solution to that
latter problem is to use a distance selection algorithm (such as in \cite{AASS}) to perform a binary search through all 
$\binom{n}{2}$ inter-point distances (the optimal $\bn$ is one of these distances), using the decision procedure to
guide the search. This yields an 
$O^*(n^{4/3})$-time algorithm for the bottleneck path problem (for points in the plane),
where the $O^*(\cdot)$ notation hides subpolynomial factors.

However, in such cases we can improve the running time, when the decision procedure has a near-linear solution,
by applying the shrink-and-bifurcate machinery of Ben Avraham et al.~\cite{BFKKS} with the recent improvements 
of Chan and Huang~\cite{CH} (see also \cite{KKSS}), to solve the bottleneck path problem in $O^*(n^{8/7})$ randomized 
expected time. See later in the paper for details.

We present two applications of the latter approach, where the major challenge is to find appropriate
near-linear implementations of the decision procedure. Typically, as already mentioned, this procedure runs BFS on
$G_{\le \bn}$. 
In typical setups, as in our applications, we are given a geometric scene consisting of a set $P$ of points in 
the plane, a range\footnote{%
  In what follows, we mostly use $r$ instead of $\bn$.}
$r = \bn > 0$, and some additional constraints, which indicate which pairs of points of $P$ can ``see'' each other 
(i.e., define edges of $G_{\le \bn}$). That is, the scene induces a \emph{visibility graph} on $P$, in which there 
is an edge between two points $p,q\in P$ if and only if they see each other, in the sense that the segment $pq$ satisfies 
the constraints imposed on the problem, and are within distance at most $r$.
For example, in one of the applications, each point $p \in P$ represents a directional antenna, i.e., a wedge $W_p$
with apex $p$, and there is an edge between $p$ and $q$ if and only if $q\in W_p$ (in the directed case) or 
$q\in W_p$ and $p\in W_q$ (in the undirected case), and the distance between $p$ and $q$ is at most $r$.

The latter approach is also suitable for the aforementioned \emph{bounded-hop} variant of the bottleneck path problem 
in $G_{\le \bn}$, where, in addition, we are given a parameter $1\le k < n$ and the goal 
is to find a path from $s$ to $t$ consisting of at most $k$ edges, with minimum bottleneck.  

Another variant of this scheme involves bottleneck paths in the visibility graph of a set $P$ of $m$ points lying on or
above a 1.5-dimensional terrain (an $x$-monotone polygonal path) with $n$ vertices in the plane. A significantly simpler 
version of this problem, without any bounded-range constraints, was recently studied in \cite{KSS}.
Here we are given a range $r > 0$, and say that two points $p,q\in P$ are mutually visible (over $T$) if
the segment $pq$ lies fully above $T$, and $|pq| \le r$. Here too, the bottleneck version receives two designated points 
$s,t\in P$, and seeks the minimum value of $r$ for which the resulting visibility graph contains a path from $s$ to $t$
(or a $k$-bounded-hop path from $s$ to $t$ in the bounded-hop version). 
In the decision version $r$ is fixed, and the goal is to determine whether the graph, with that $r$, contains a path
from $s$ to $t$ (or a path with at most $k$ edges).

\subparagraph*{Related work.}
Breadth-First Search (BFS) is recognized as one of the fundamental graph algorithms. In general, its running time 
is $O(|V|+|E|)$, which is inefficient when $|E|$ is large. Its importance has led authors to seek families of graphs
in which BFS can be implemented more efficiently, i.e., in subquadratic time that only depends on the number
of vertices. One such example is the family of disk graphs. The vertices of a
disk graph represent disks in the plane and there exists an edge between two vertices if and only if the
corresponding disks intersect. Notice that the number of edges in a disk graph can be quadratic in $n$, the number of
disks. Nevertheless, for unit-disk graphs, Cabello and Jej\v{c}i\v{c}~\cite{CabelloJ15} presented an
$O(n \log n)$ implementation of BFS, and subsequently Chan and Skrepetos~\cite{CS} presented an alternative $O(n)$ implementation (after pre-sorting the points by their $x$- and $y$-coordinates). For arbitrary disks, Kaplan et al.~\cite{KKSS} and Liu~\cite{Liu} described an $O(n \log^4 n)$ implementation of BFS, which was recently improved to $O(n \log^2 n)$ by Klost~\cite{Klost23} and subsequently to $O(n \log n)$ by Cabello and de Berg~\cite{BergC25}. 

The proximity graph of a set $S$ of $n$ segments in the plane, for a real parameter $r \ge 0$, is 
$G_r(S) \coloneqq (S,E)$, where $E = \{ (e_1,e_2) \mid \dist(e_1,e_2) \le r\}$
 and $\dist (e_1,e_2)$ is the Euclidean distance between $e_1$ and $e_2$ (which is 0 if they intersect). Agarwal et
al.~\cite{AgarwalKS24} devised an $O^*(n^{4/3})$ implementation of BFS in $G_r(S)$. For the special case where the
segments in $S$ are pairwise disjoint, Agarwal et al.~\cite{AgarwalKKS24} have later provided an $O(n \log^2 n)$ 
implementation.

The bottleneck path problem (also known as the minimax path problem) and its complementary problem, i.e., the widest
path problem (or the maximum capacity problem), are well-known problems in graph theory. Abu-Affash et
al.~\cite{Abu-AffashCKS14} studied the problem of placing at most $k$ Steiner points to minimize the bottleneck of a
path between two designated points in the plane, for a given integer parameter $k \ge 0$, and developed an $O(n
\log^2 n)$-time algorithm for the problem.

The reverse shortest path (RSP) problem for unit disks (mentioned above) was studied by Wang and Zhao~\cite{WZ}, who
proposed an $O^*(n^{5/4})$-time solution. An improved solution with running time $O^*(n^{6/5})$, using the
shrink-and-bifurcate technique, was subsequently presented by Kaplan et al.~\cite{KKSS}. A recent improvement
of the shrink-and-bifurcate technique, due to Chan and Huang~\cite{CH}, allows one to improve the running time
to $O^*(n^{9/8})$. The RSP problem was also
studied for other objects, including wedges of some fixed angle, which are viewed as directional antennas. This
latter version, which is especially relevant to our current work, was solved in $O^*(n^{4/3})$ time by 
Agarwal et al.~\cite{AgarwalKS24}.  

Finally, wireless networks, and in particular those utilizing directional antennas, as defined above, have received
significant attention in recent years; see, e.g., the series of papers dealing with the bounded-angle spanning 
tree problem~\cite{AschnerK17,AshurK23,BiniazBLM22,BiniazDM23}. 

\subparagraph*{Our results.}
Our main contribution is in providing near-linear implementations of BFS in two types of visibility graphs arising
in realistic scenarios. It is interesting and somewhat unexpected that such implementations exist, since at first sight it seems
that employing simplex range searching, which would immediately increase the running time to at least $O^*(n^{4/3})$,
is inevitable, when the constraints defining the graph involve halfplane or disk range searching, as 
in our applications, or in more general setups. 

Let $P$ be a set of $n$ points in the plane, where each point $p \in P$ has a wedge $W_p$ with apex at $p$ associated
with it. Let $\alpha > 0$ be such that each wedge is of angle at least $\alpha$. (This is a key assumption, without which a running time near $n^{4/3}$ seems inevitable.) Let $r > 0$ be a real parameter. The wedges,
viewed as directional antennas of range $r$, naturally induce two types of graphs. The first is the directed
visibility graph $\overrightarrow{VG}(P,r)$ of $P$, where there is a directed edge from $p \in P$  to $q \in P$ if
and only if $q \in W_p$ and $|pq| \le r$. The second type is the undirected visibility graph $VG(P,r)$, where there
is an edge between $p$ and $q$ if and only if $p \in W_q$, $q \in W_p$, and $|pq| \le r$. In
Section~\ref{sec:antennas-bfs} we devise near-linear implementations of BFS in these graphs. We then apply the
shrink-and-bifurcate technique to solve the (general or bounded-hop) bottleneck path problem in these graphs in
$O^*(n^{8/7})$ randomized expected time, using BFS as the decision procedure. The $O^*(n^{8/7})$ bound improves the
$O^*(n^{4/3})$ bound of Agarwal et al.~\cite{AgarwalKS24}, mentioned earlier, which holds in a slightly more general
setting. 

We also study bounded-range visibility graphs of points on or above a $1.5$-dimensional terrain $T$ (i.e., an 
$x$-monotone polygonal curve). That is, there is an edge between two points $p$ and $q$ if and only if the line 
segment $pq$ lies above $T$ and  $|pq| \le r$. Recently, Katz et al.~\cite{KSS} presented a near-linear 
implementation of BFS in such graphs without the bounded-range constraint (i.e., assuming $r = \infty$). In 
Section~\ref{sec:ter}, we present a near-linear implementation for any finite $r$, which is a significantly more
difficult problem. Here too, by combining this procedure with the shrink-and-bifurcate technique, we solve the 
corresponding bottleneck path problem in $O^*(n^{8/7})$ randomized expected time.

\section{The case of antennas} \label{sec:antennas-bfs}

\subsection{BFS in directed visibility graphs induced by directional antennas} \label{subsec:antennas-bfs}

\begin{figure}[ht]
\centering
\includegraphics[scale=1]{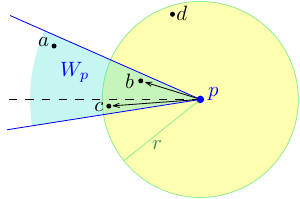}
\caption{The points $b$ and $c$ lie in $W_p$ and are at distance at most $r$ from $p$. The point $a$ is out of range,
and $d$ is not in $W_p$.}
\label{fig:dhv}
\end{figure}

Let $P$ be a set of $n$ points in the plane, each representing a \emph{directional antenna} of angle at least
$\alpha$, for some $\alpha >0$. Specifically, each point $p \in P$ has a wedge $W_p$ associated with it, 
whose apex is $p$, with an opening angle at least $\alpha$. 
Let $r$ be a real parameter and let $VG(P,r)$ be the directed visibility graph of $P$, where there is a directed edge 
$(p,q)$ from $p \in P$ to $q \in P$ if and only if $q\in W_p$ and the distance between $p$ and $q$ is at most $r$ (in 
sensor-network parlance, $q$ is within the transmission range of $p$); see Figure~\ref{fig:dhv}. In this section we present 
a near-linear algorithm for performing BFS in the graph $VG(P,r)$, from a given start point $s \in P$. 
This is nearly the best algorithm one can hope for, recalling that the graph may have quadratically many edges.

The general scheme goes as follows. As usual, we construct the BFS layer by layer; the $i$-th layer is denoted by $L_i$, 
starting with $L_0 = \{s\}$. Suppose we have already constructed $L_i$ and now want to construct $L_{i+1}$. 
We maintain the set of points of $P$ that the BFS has not reached yet; denote by $P_i$ this set at the end 
of the $i$-th iteration.

We construct and maintain the following dynamic data structure on $P_i$. We query it with the points of $L_i$, where the goal
of the queries is to find and report all the points $p\in P_i$ that are visible from at least one point of $L_i$ (in the 
sense that there exists $q\in L_i$ such that $p\in W_q$ and $|pq|\le r$), and delete them from the structure as soon
as they are discovered, to ensure that no point is reported more than once. By definition, these points comprise $L_{i+1}$.
Their deletion from $P_i$ yields $P_{i+1}$.

Dropping the index $i$ for convenience, the setup is thus as follows.
We have a set $P$ of points in the plane,  which we want to maintain under deletions, so as to be able 
to answer efficiently queries, each of which specifies a query point $q$ (representing an antenna), and seeks
to report all the points of $P$ that are visible from $q$ (i.e., lie in $W_q$), and at distance at most $r$ from it.
The problem we face is that the condition for $qp$ to be an edge is a conjunction of three constraints:
$p$ has to (i) lie counterclockwise to the clockwise ray bounding $W_q$, (ii) lie clockwise to the 
counterclockwise ray bounding $W_q$, and (iii) be at distance $\le r$ from $q$.
A na\"ive, albeit inefficient, implementation would use a three-level (dynamic) reporting data structure,
which would result in a significantly superlinear algorithm. To avoid this inefficiency, we adopt
the following approach.
The idea is to partition the plane into axis-aligned regions, which induce a partition of $P$ into subsets. 
We then show that for each such subset it is sufficient to check only one of the above three constraints.
That is, in some of the subsets it suffices to report all points at distance at most $r$ from $q$ 
(using a dynamic Voronoi diagram), in other subsets it suffices to report all points that are 
counterclockwise to the clockwise ray of $W_q$, and in other subsets it suffices to report all 
points that are clockwise to the counterclockwise ray of $W_q$
(using a dynamic halfplane range reporting data structure in each of the two latter cases).
In each of these three cases, the other two constraints are automatically satisfied. 

For this, we prepare the following (dynamic) multi-level data structure for the subsets of $P$
in each of the regions. 
We first construct a two-level dynamic orthogonal range tree,
where the primary tree $T$ is constructed on the $y$-order of the points of $P$, and at each node $v$ of $T$, we construct a 
secondary tree $T_v$ on the points stored at the subset of $T$ rooted at $v$, sorted by their $x$-coordinates.
Finally, at each node $\eta$ of each $T_v$, we create two dynamic third-level structures on the subset $P_{v,\eta}$ of 
the points stored at the 
subtree rooted at $\eta$. One structure is a dynamic Voronoi diagram of $P_{v,\eta}$, and the other is a dynamic halfplane
range reporting structure for $P_{v,\eta}$. The Voronoi diagram is constructed and maintained using the technique of
Kaplan et al.~\cite{KMRSS}, improved by Liu~\cite{Liu}. The halfplane range reporting structure is constructed and 
maintained using the dynamic algorithm of Overmars and van Leeuwen~\cite{OvL}. In the latter structure, a deletion
costs $O(\log^2n)$ time, and a query takes $O(\log n)$ time, which are dominated by the $O(\log^4n)$ time for searching 
and updating the Voronoi diagram. We note that these running time bounds for deletions are amortized.

\begin{figure}[ht]
\centering
\includegraphics[scale=0.8]{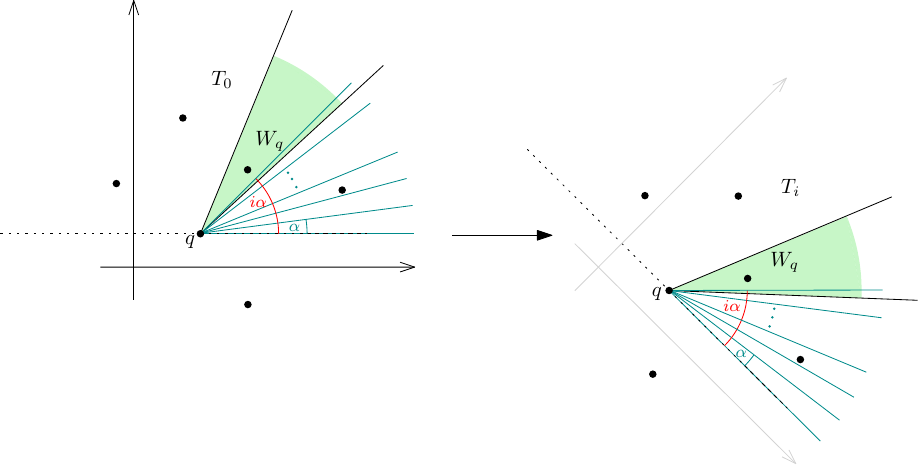}
\caption{Left: The original scene and a query wedge $W_q$. Right: The query will be performed in the $i$'th copy
$T_i$, after rotating $W_q$ clockwise by $i\alpha$ degrees.}
\label{fig:verRot}
\end{figure}

Finally, we maintain $\lceil \frac{\pi}{\alpha} \rceil$ copies of the data structure,
denoted $T_0, T_1, T_2, \ldots$, where $T_i$ stores a copy of $P$ obtained by rotating the
original scene clockwise about the origin by the angle $i\alpha$, so that lines of (original) orientation $i\alpha$
become horizontal. Given a query point $q$ associated with a wedge $W_q$, we perform the query in the copy $T_i$,
where $i$ is the smallest index for which $W_q$ is intersected by the line through $q$ at orientation $i\alpha$
(actually, any index $i$ with this property will do). To
perform the query in $T_i$, we rotate the wedge $W_q$ clockwise by $i\alpha$ degrees, so that $W_q$ contains either
the rightward or the leftward horizontal ray emanating from $q$. See Figure~\ref{fig:verRot}. 

\begin{figure}[ht]
\centering
\includegraphics[scale=0.8]{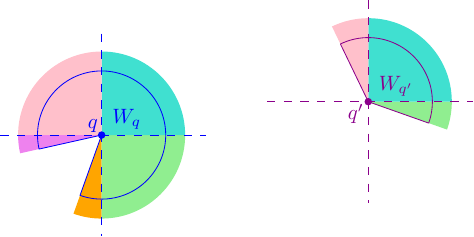}
\caption{Horizontal and vertical lines through a point divide its wedge into at most five
sub-wedges, each of angle at most $\pi/2$ and with at least one axis-parallel ray.}
\label{fig:wedgPart}
\end{figure}

For convenience, we refer to the chosen data substructure as $T$.
Given a query point $q$ with associated wedge $W_q$ that contains a horizontal ray emanating from $q$, we divide
$W_q$ by horizontal and vertical lines through $q$, into at most five sub-wedges, each of angle at most
$\pi/2$ and with at least one axis-parallel ray; see Figure~\ref{fig:wedgPart}. 
We perform the query on each sub-wedge separately. Without loss of generality, we present here 
the query process for a sub-wedge whose lower ray is a horizontal right ray.

\begin{figure}[ht]
\centering
\includegraphics[scale=0.8]{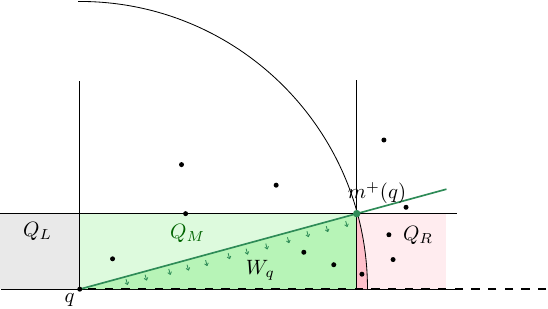}
\caption{The rectangular regions arising when querying with a point $q$, for the points above $q$. $m^+(q)$ is the point 
at distance $r$ from $q$ along the upper ray of $W_q$. Points in $P_R = P\cap Q_R$ lie in $W_q$, and points in 
$P_M = P\cap Q_M$ are at distance at most $r$ from $q$. A symmetric partition holds for the points below $q$.}
\label{fig:dhr}
\end{figure}

For convenience, we refer to this sub-wedge also as $W_q$. Let $m^+(q)$ be the point at distance 
$r$ from $q$ along the upper ray of $W_q$; see Figure~\ref{fig:dhr}. We first search in the first level of $T$ with 
(the $y$-coordinates of) $q$ and $m^+(q)$. This yields all the points of $P$ lying in the slab $\sigma$ bounded by the
horizontal lines through $q$ and $m^+(q)$, as the disjoint union of $O(\log n)$ canonical sets, each being the subset stored 
at some subtree of $T$. At each such subtree, rooted at some node $v$, we search in the secondary tree $T_v$ with the 
$x$-coordinates of $q$ and $m^+(q)$, thereby splitting $\sigma$ into three rectangular regions, a semi-unbounded region $Q_L$
lying to the left of $q$, a bounded rectangle $Q_M$ lying between $q$ and $m^+(q)$, and another semi-unbounded region 
$Q_R$ lying to the right of $m^+(q)$. Notice that in case $W_q$ is of angle $\pi/2$, $Q_M$ is empty. Put $P_L = P\cap
Q_L$, $P_M = P\cap Q_M$, and $P_R = P\cap Q_R$. Clearly, no point of
$P_L$ can lie in $W_q$. 

For $P_R$, all its points lie in $W_q$, so it suffices to report those points at distance $\le r$ from $q$. To do so, 
we query with $q$ at the corresponding dynamic 
Voronoi diagrams of the $O(\log^2n)$ canonical subsets $P_{v,\eta}$ that comprise
$P_R$. Each output point $p$ at distance $\le r$ from $q$ is added to $L_{i+1}$, and is promptly deleted from all 
the $O(\log^2n)$ structures (Voronoi diagrams and dynamic halfplane range reporting structures) that it participates in.
We stop querying with $q$ when no neighbor at that distance is found.

For $P_M$, all its points are at distance $\le r$ from $q$, so it suffices to report those points that lie in $W_q$.
By construction, this is equivalent to reporting all the points of $P_M$ that lie below the upper ray of $W_q$.
We follow the same approach as before, but now we query at the halfplane range reporting structures corresponding to the
relevant canonical subsets.

Note that no point above the horizontal line through $m^+(q)$ is visible from $q$ --- it either is too far from $q$ 
or lies above $W_q$. See Figure~\ref{fig:dhr} for an illustration. 

Symmetric procedures are applied for each of the other cases---sub-wedges below the horizontal line through $q$, 
sub-wedges for which the axis-parallel bounding ray is vertical, and so on.

As there are $O(\log^2n)$ canonical subsets that the query processes, the overall cost of a query (i.e., the part 
of the query that finds a neighbor of $q$ or determines that no such neighbor exists), followed by deletion
when a neighbor is found, is $O(\log^6n)$ (which is amortized for deletions). 

The correctness of the query procedure is straightforward, and follows from the arguments given above. 

We now run BFS on $VG(P,r)$, using the above data structure, in the manner described earlier. Each point of $P$ acts as a query
only at the layer it belongs to. The cost of querying with a point $q$ is $O((1+k_q)\log^6n)$, where $k_q$ is the number 
of (yet unvisited) neighbors of $q$ that the algorithm detects. Since no point is reported more than once (since it is deleted from all 
its containing structures once it is discovered), we have $\sum_q k_q \le n$, so the overall cost of the algorithm is 
$O(n\log^6n)$. That is, we have shown:
\begin{theorem} \label{thm:bfs1}
BFS on $VG(P,r)$ can be performed in $O(n\log^6n)$ time.    
\end{theorem}

\subsection{The corresponding bottleneck path problem}

Recall that the bottleneck path problem is to find the minimum value $r^*$ for which there exists a path from $s$ to $t$ 
in $VG(P,r^*)$. This optimization problem can be solved using the shrink-and-bifurcate technique of \cite{BFKKS,KKSS,CH}, 
as mentioned in the introduction, using the BFS algorithm of Theorem~\ref{thm:bfs1} as the decision procedure.

\subparagraph*{A brief review of the shrink-and bifurcate technique.}
In more details, this technique is applied when we have an efficient decision procedure, for which we do not have 
an efficient parallel implementation to facilitate standard parametric search. What we do is first run an 
interval-shrinking phase, which produces an interval $I$ that contains a small number of critical values, including 
the optimal value $r^*$. We then simulate the decision procedure sequentially, where most of the comparisons 
(those whose critical values lie outside $I$) can be resolved. For those comparisons with critical values in $I$,
we bifurcate, following both outcomes of the comparison. We stop the bifurcation when we exceed a certain threshold
number of them, or when the simulation has executed at least some threshold number of steps,
and then resolve all of them using the standard binary search employed by parametric search. Then
a new bifurcation stage begins, at the leaf found to contain $r^*$,
and we proceed this way until we simulate the entire decision procedure.

A careful choice of parameters makes this procedure efficient. For example, when the critical values 
are inter-point distances in the plane, this shrink-and-bifurcate procedure can be implemented to run 
in $O^*(n^{8/7})$ randomized expected time~\cite{CH}.

This sketchy review only provides some highlights of the technique. See \cite{BFKKS} and \cite{KSS} 
for full details, and see \cite{CH} for a recent improved implementation of the shrinking phase. 

\subparagraph*{Applying the shrink-and-bifurcate technique.} 
In our context, the critical values of $r$, namely values at which the outcome to some comparison
made by the BFS changes, are of two kinds: (i) pairwise distances between the points of $P$, and
(ii) values at which some point $m^+(q)$ (or its symmetric counterpart along the lower ray of $W_q$) 
becomes co-vertical or co-horizontal with some point of $P$. We handle each kind of critical values 
separately. We begin the procedure by sorting all the points of $P$ into the two-dimensional
orthogonal range tree $T$. We then locate the $y$-coordinates of the points $m^+(q)$ in the primary 
tree, to obtain the canonical subsets that represent the points below and above each $m^+(q)$.

The simulation of this step, with the unknown $r^*$, can be performed using standard parametric search, since we
can search with the points $m^+(q)$ in parallel, and each search only takes $O(\log n)$ steps.

Similarly, for each canonical set $T_v$ that we obtain for some $m^+(q)$, we search with the $x$-coordinate of
$m^+(q)$ in $T_v$, to obtain a representation of the points of $T_v$ to the left and to the right of $m^+(q)$. 
This too can be done in parallel, over all points $q$, all nodes $v$, and the $O(\log n)$ steps of 
each binary search.

A symmetric procedure is applied to the corresponding points on the lower rays.

Hence, when this part of the simulation ends, all critical values of type (ii) are out of the interval that contains
$r^*$, as obtained by the standard parametric search technique, and each (future) comparison that involves 
them can be resolved, independently of $r^*$.

For critical values of type (i), we use the shrink-and-bifurcate technique of \cite{BFKKS,KKSS} reviewed above.
Using the implementation of Chan and Huang~\cite{CH} for the shrinking part of the procedure yields 
a procedure that runs in $O^*(n^{8/7})$ randomized expected time. 

Note that the same technique can also solve, with the same performance bounds, the aforementioned 
bounded-hop or RSP (reverse shortest path) version of the problem,
in which we are given an additional integer parameter $k$, and seek the minimum $r^*$ for which $VG(P,r^*)$ contains
a path from $s$ to $t$ of length at most $k$. That is, we have:

\begin{theorem}
The bottleneck path problem, as well as its bounded-hop (RSP) version, for directed visibility graphs 
induced by $n$ directional antennas in the plane, each of opening angle at least $\alpha$, for some
constant parameter $\alpha>0$, can be solved in $O^*\left(\frac{1}{\alpha} n^{8/7}\right)$ 
randomized expected time.
\end{theorem}

\subsection{BFS in undirected visibility graphs induced by directional antennas} \label{subsec:antennas-bfs-undir}

Let $r, \alpha$ be positive real parameters and let $P$ be a set of $n$ points in the plane, each representing 
a directional antenna, with an associated wedge $W_p$ of angle $\ge\alpha$. We consider here the 
setup, in which two antennas, with apices $p$, $q$, can see (or rather communicate with) each other
if and only if $q\in W_p$, $p\in W_q$, and $|pq|\le r$. The corresponding undirected graph $VG(P,r)$ 
consists of all mutually visible pairs $(p,q)$. Given two antennas 
$s,t\in P$, the goal is to determine whether $t$ is reachable from $s$ in $VG(P,r)$, or, in the bounded-hop 
version, whether $t$ is reachable from $s$ along a path with at most $k$ edges, for an additional parameter $k$. 

To solve the reachability problem, we use a similar scheme to that in the directed version, 
modifying only the query process. Following the procedure described in Section~\ref{subsec:antennas-bfs}, 
without loss of generality, we present the query process for a query point $q$ and its sub-wedge of angle 
at most $\pi/2$ whose lower ray is a horizontal right ray. That is, we detect
only the visible points lying above and to the right of $q$. As in the directed version, a symmetric process is
applied to each of the other sub-wedges of $W_q$.

\begin{figure}[ht]
\centering
\includegraphics[scale=0.7]{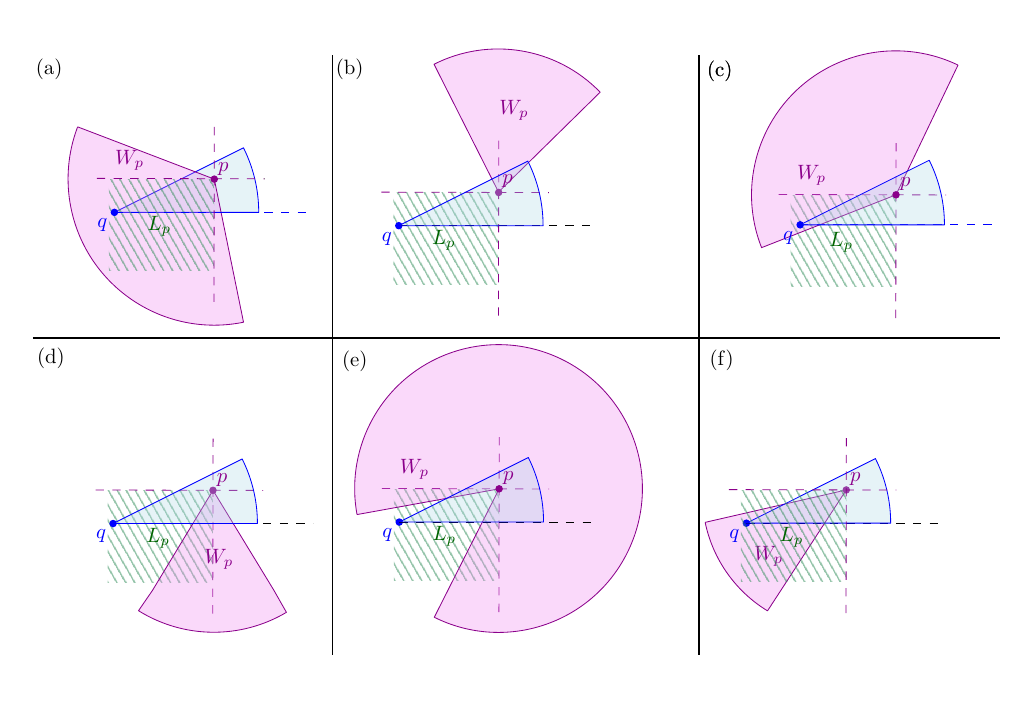}
\caption{The query in the undirected case. (a) $L_p\subseteq W_p$, i.e., $q\in W_p$. 
(b) $L_p\cap W_p=\emptyset$, in which case $p$ and $q$ do not see each other. 
(c) $L_p\cap W_p$ is a sub-wedge that contains the left-horizontal ray from $p$. 
(d) $L_p\cap W_p$ is a sub-wedge that contains the lower-vertical ray from $p$.
(e) $L_p\cap W_p$ is the union of two disjoint sub-wedges, one of type (c) and one of type (d).
(f) $W_p\subset L_p$, so that it does not contain any axis-parallel ray from $p$. }
\label{fig:UndrctWdgOpts}
\end{figure}

Notice that for each antenna $p$ lying above and to the right of $q$, the point $q$ lies in the lower-left quadrant
of $p$. Therefore, we handle antennas in different ways, depending on how their wedge intersects this lower-left 
quadrant. See Figure~\ref{fig:UndrctWdgOpts}, where the lower-left quadrant of $p$ is denoted by $L_p$.

\begin{itemize}
    \item For points $p$ with $L_p \subseteq W_p$---guaranteeing that $q \in W_p$ 
(Figure~\ref{fig:UndrctWdgOpts}(a))---it suffices to detect every $p$ lying in $W_q$ at distance at most 
$r$ from $q$. This coincides with the visibility definition in the directed version, and 
we simply follow the corresponding query process described there.
    \item If $L_p \cap W_p = \emptyset$ (Figure~\ref{fig:UndrctWdgOpts}(b)), we can safely conclude that $q \notin
W_p$, and therefore $p$ and $q$ are not visible to each other.
    \item The case where $W_p$ contains only the horizontal ray bounding $L_p$ but not the vertical one 
is referred to as {\bf case (c)} in Figure~\ref{fig:UndrctWdgOpts}, and is studied in 
Section~\ref{subsubsec:case_c}.
    \item The case where $W_p$ contains only the vertical ray bounding $L_p$, but not the horizontal 
one, referred to as {\bf case (d)} in Figure~\ref{fig:UndrctWdgOpts}, is symmetric to the former 
case (c).
    \item The case where $W_p \cap L_p$ is the union of two disjoint sub-wedges 
(Figure~\ref{fig:UndrctWdgOpts}$(e)$), one of type (c) and the other of type (d), 
is handled by applying the corresponding procedure to each sub-wedge separately.
    \item Finally, consider the points $p$ with $W_p\subset L_p$, where $W_p$ contains 
no axis-aligned rays, see Figure~\ref{fig:UndrctWdgOpts}$(f)$.
In Section~\ref{subsubsec:case_f}
we present a further multi-level partition of these points into $\frac{\pi}{2\alpha}$ subsets, 
and each subset is split into points that can be handled as in case (c), and points that can be 
handled as in case (d).  
\end{itemize}


\subsubsection{Handling case (c)} \label{subsubsec:case_c}
\begin{figure}[ht]
\centering
\includegraphics[scale=0.7]{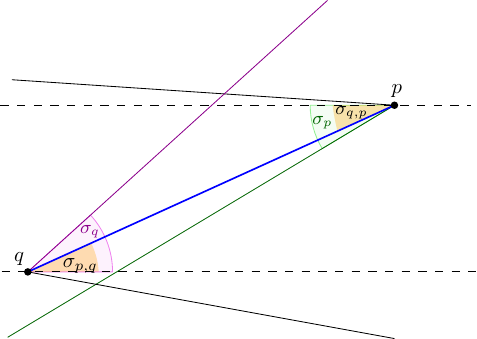}
\caption{For $p$ lying above $q$, $p$ and $q$ lie in each other's wedge if and only if $p$ is below the upper ray 
of $W_q$ and $q$ is above the lower ray of $W_p$. Moreover, if $\sigma_p \le \sigma_q$, as in the figure, 
then for $p$ and $q$ to lie in each other's wedge it suffices that $q$ be above the lower ray of $W_p$. A symmetric single condition holds when $\sigma_p \ge \sigma_q$.}
\label{fig:hrt}
\end{figure}

In this case $p$ lies above the lower ray of $W_q$, and $q$ lies below the upper ray of $W_p$,
so it suffices to test the position of $p$ with respect to the upper ray of $W_q$, and the position of $q$ with
respect to the lower ray of $W_p$. Let $\sigma$ be the counterclockwise angle between the rightward-directed 
horizontal ray emanating from $q$ and $qp$ (the same as between the leftward-directed horizontal ray emanating 
from $p$ and $pq$), see Figure~\ref{fig:hrt}.

Denote by $\sigma_p$ (resp., $\sigma_q$) the angle of the sub-wedge at $p$ (resp., at $q$)
between its lower ray (resp., upper ray) and the $x$-axis.
Observe that $p$ lies below the upper ray of $W_q$ if and only if $\sigma \le \sigma_q$, and 
$q$ lies above the lower ray of $W_p$ if and only if $\sigma \le \sigma_p$. 
It therefore suffices to test whether $\sigma \le \min\{\sigma_q,\sigma_p\}$.

We therefore distinguish between two main cases, and handle each of them separately.
First, consider those $p$ with $\sigma_q \le \sigma_p$.
In this case if $p$ lies below the upper ray of $W_q$, then $q$ certainly lies
above the lower ray of $W_p$, hence we report all the points that lie below the upper ray of $W_q$, and are at
distance at most $r$ of $q$. This is done exactly as in Section~\ref{subsec:antennas-bfs}; see Figure~\ref{fig:dhr}. 

\begin{figure}[ht]
\centering
\includegraphics[scale=0.7]{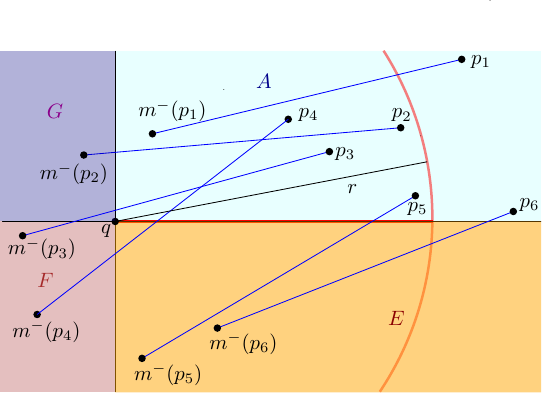}
\caption{Partition of the plane into regions. Points $p \in A$ and the corresponding segments $pm^-(p)$.}
\label{fig:regions}
\end{figure}

Next, consider those $p$ with $\sigma_p \le \sigma_q$ (this is the case depicted in Figure~\ref{fig:hrt}). 
Here we need to report all the points $p$ at distance at most $r$ from $q$, such that $q$ lies above the lower ray of $W_p$
(and then $p$ certainly lies below the upper ray of $W_q$). This case is more involved and is handled as follows.
Consider the partition of the plane into four quadrants obtained by drawing a horizontal and a vertical line through
$q$. We denote the quadrants starting from the northeast one and advancing clockwise by $A$, $E$, $F$, and $G$,
respectively; see Figure~\ref{fig:regions}. Recall that we are only interested in points $p$ that lie above and to
the right of $q$, i.e., in $A$.

Let $m^-(p)$ be the point that lies at distance $r$ from $p$ along the lower ray of $W_p$, and
distinguish between several cases, depending on which of the four quadrants $A, E, F, G$ defined by $q$ contains $m^-(p)$. 

If $m^-(p)$ lies in the southeast quadrant $E$ (see points $p_5$ and $p_6$ in Figure~\ref{fig:regions}), then clearly
$q$ lies above the lower ray of $W_p$, and it suffices to test whether $|pq|\le r$, 
by searching in the suitable Voronoi diagrams.

If $m^-(p)$ lies in the northwest quadrant $G$ (see $p_2$ in Figure~\ref{fig:regions}) then clearly $q$ lies below
the lower ray of $W_p$, so this case can be ignored.  

Consider the case where $m^-(p)$ lies in the southwest quadrant $F$ (see $p_3$ and $p_4$ in
Figure~\ref{fig:regions}). Then both segments $m^-(p)q$ and $qp$ have
positive slopes, implying that the angle $\angle m^-(p)qp$ between them is obtuse. Therefore, $|pq| < |pm^-(p)| = r$,
so it suffices to test whether $q$ lies above the lower ray of $W_p$.
(This step is implemented in the 
dual plane, as a halfplane range reporting query with the line dual to $q$ amid the points dual to the lines containing the 
lower rays of the appropriate set of yet undiscovered antennas with a leftward-downward directed lower ray.)

Finally, consider the case where $m^-(p)$ lies in the northeast quadrant $A$ (see $p_1$ in 
Figure~\ref{fig:regions}). Here both segments $qm^-(p)$ and $m^-(p)p$ have positive slopes, 
implying that the angle $\angle qm^-(p)p$ between them is obtuse. Therefore, 
$|pq| > |pm^-(p)| = r$, so this case can be ignored too.

To recap, knowing the quadrant containing $m^-(p)$ allows us either to discard $p$ altogether as a possible 
neighbor of $q$, or to determine which of the two reporting tasks (halfplane range reporting in the dual 
plane or searching in Voronoi diagrams) suffices to determine whether $p$ is a neighbor of $q$.

\begin{figure}[ht]
\centering
\includegraphics[scale=0.7]{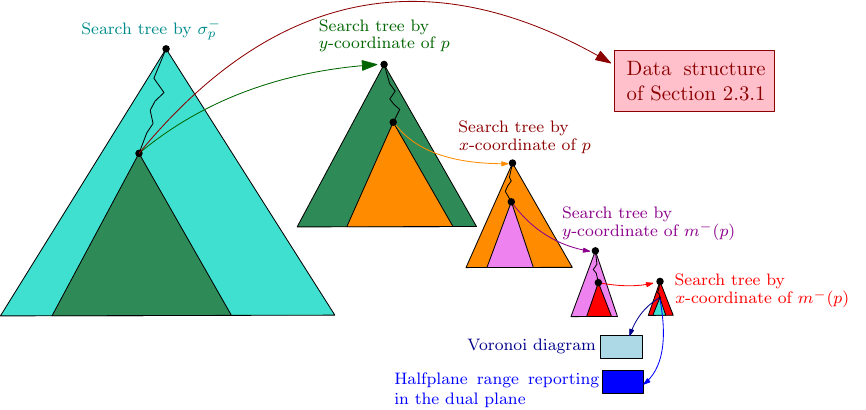}
\caption{A schematic illustration of the data structure used in the undirected visibility graph for case (c).}
\label{fig:dataSt}
\end{figure}

The actual data structure is constructed, and searched in, as follows. 
For concreteness, and without loss of generality, we continue to focus on the task of 
reporting the points $p$ that lie above and to the right of the query point $q$,
and satisfy $p \in W_q$, $q \in W_p$, and $|pq| \le r$. The tasks of reporting the points 
of $P$ in the other three cases are treated in a fully symmetric manner. We construct an 
orthogonal range tree $T_\sigma$ on the angles $\sigma^-_p$, for $p$ in the set of antennas 
being maintained, where $\sigma^-_p$ (formerly referred to as $\sigma_p$) is the angle between 
the left horizontal direction and the lower ray of $W_p$. We search in $T_\sigma$ with the angle 
$\sigma^+_q$, the angle between the right horizontal direction and the upper ray of $W_q$,
to obtain a collection of $O(\log n)$ canonical subsets, in each of which we know whether the 
angles $\sigma_p^-$ of the points $p$ in the subset are all smaller or all larger than $\sigma_q^+$.
Consequently, for each of the canonical subsets of $T_\sigma$, we construct two data structures, 
one for each of the two cases $\sigma^-_p \le \sigma^+_q$ and $\sigma^-_p \ge \sigma^+_q$, for
all $p$ in the subset. In order to deal with the latter case, we construct and use, at each 
canonical set stored at some node of $T_\sigma$, the data structure of Section~\ref{subsec:antennas-bfs}. 
In order to deal with the former case, we construct the following multilevel data structure. 
The first four levels of the structure constitute a four-dimensional orthogonal range tree, 
where the first two levels are constructed on the $y$-coordinates and $x$-coordinates of the 
points $p$, and the last two levels are constructed on the $y$-coordinates and $x$-coordinates of
the points $m^-(p)$. Again, this is done for each canonical set stored at some node of the top 
level $T_\sigma$. By searching with $q$ in these four-level range trees, we determine the points 
$p$ above and to right of $q$, whose associated points $m^-(p)$ lie in each of the two lower 
quadrants determined by $q$. (By the preceding analysis we can ignore the two upper quadrants.) 
For the canonical subsets of points $p$ whose associated points $m^-(p)$ lie in the southeast 
quadrant $E$ determined by $q$, we maintain a dynamic Voronoi diagram and use it to detect and
report all the points at distance at most $r$ from $q$. For those canonical subsets of points 
$p$ whose associated points $m^-(p)$ lie in the southwest quadrant $F$, we maintain a dynamic
halfplane range reporting structure in the dual plane, to detect and report all the points $p$ 
for which $q$ lies above the lower ray of their wedge $W_p$. See Figure~\ref{fig:dataSt} for an 
illustration of the structure. 

\old{
\subsubsection{Handling case (d)} \label{subsubsec:case_d}
This case is similar to case (c) and is differed to Appendix~\ref{app:case_d}. 
}

\old{
Here we are given a set of points $p$ such that the lower ray of $W_q$ lies below $p$ (more
precisely, the clockwise ray of $W_q$ lies clockwise to $\vec{qp}$), and the lower ray of $W_p$
lies counterclockwise to $q$. Our goal is to detect all points $p$ such that $q$ lies below 
the upper ray of $W_p$, $p$ lies below the upper ray of $W_q$, and $|pq|\le r$.

Let $H$ be a vertical line between $q$ and this set of points. By Lemma 1 in \cite{KSS},
we know the following: for any $p$ where the upper ray of $W_p$ intersects $H$ above upper ray
of $W_q$, if $p$ lies below the upper ray of $W_q$, then $q$ must lie below the upper ray of $W_p$. 
Similarly, for any $p$ where the upper ray of $W_p$ intersects $H$ below the upper ray of $W_q$, 
if $q$ lies below the upper ray of $W_p$, then $p$ must lie below
the upper ray of $W_q$ (see Figure 3(b) in \cite{KSS}). 

\micha{I did not process the rest of this subsection. In particular, I don't understand why case (d) has to be treated differently than case (c). If there is a real reason for this, it should be explained in a sentence or two.}

Putting this together, for points $p$ where the upper ray of $W_q$ intersects $H$ below  the upper ray of $W_p$, we
detect all points $p$ lying below the upper ray of $W_q$ with $|pq|\le r$. This is done exactly as in
Section~\ref{subsec:antennas-bfs}; see Figure~\ref{fig:dhr}. 
Symmetrically, for points $p$ where the upper ray of $W_p$ intersects $H$ below the upper ray of $W_q$, we detect all
points $p$ such that $q$ lies below the upper ray of $W_p$ and $|pq|\le r$. This is achieved similarly to
Section~\ref{subsubsec:case_c}, by considering the possible locations of the points $m(p)$ on the upper ray of $W_p$
at distance $r$ from $p$, with respect to the four quadrants around $q$. As analyzed in
Section~\ref{subsubsec:case_c}, if $m(p)\in A$, then $r = |pm(p)|<|pq|$, if $m(p)\in F$ then $|pq|\leq |pm(p)| = r$,
if $m(p)\in G$ then $q$ below the ray contains $pm(p)$, and if $m(p)\in E$ then $q$ above the ray contains $pm(p)$
(see Figure~\ref{fig:regions}). Here, however, we consider points $m(p)$ lying on upper rays, rather than lower rays,
hence the roles for quadrants $F$ and $G$ swapped. All together, all points $p$ with $m(p)\in E\cup A$ can be
ignored, for all the points $p$ with $m(p)\in G$, we detect those at distance at most $r$ from $q$, and for all the
points $p$ with $m(p)\in F$ we detect those with upper ray that lie above $q$.
}


\subsubsection{Handling case (f): $W_p\subset L_p$} \label{subsubsec:case_f}

Here we are given a set of points $p$ with $W_p \subset L_p$, where $W_p$ does not contain the horizontal and vertical rays of $L_p$.
We partition the set into subsets of points where either all have their upper ray above $q$
(handled as in case $(c)$) or all have their lower ray below $q$ (handled as in case $(d)$). 
We achieve this by first
dividing the set into $\ell = \lceil \frac{\pi}{2\alpha} \rceil$ subsets, where the $i$-th subset, $i=1,\ldots,\ell$, consists of
all points $p$ for which $i$ is the minimal index such that $W_p$ contains a ray in direction $d_i = \pi + i\alpha$
(recall that all wedges have angle at least $\alpha$ and each has an upper ray in a direction
greater than $\pi$).

\begin{figure}[ht]
\centering
\includegraphics[scale=1.0]{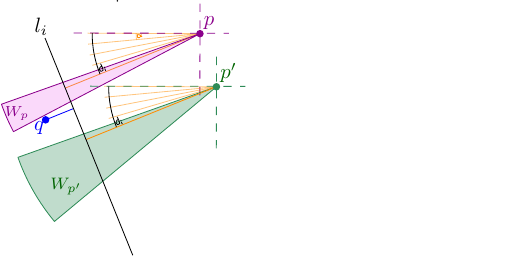}
\caption{Both $p$ and $p'$ belong to the $i$-th subset, and $l_i$ is the line perpendicular to $d_i$. $q$'s projection onto $l_i$ is below that of $p$, hence $q$ surely lies
below the upper ray of $W_p$. Symmetrically, $q$ surely lies above the lower ray of $W_{p'}$.}
\label{fig:rotateF}
\end{figure}

We store the $i$-th subset in a balanced binary search tree, ordered by the projections of its points onto $l_i$, the
line perpendicular to $d_i$. Notice that if the projection of $p$ onto $l_i$ is higher than that of $q$, we can
safely conclude that $q$ lies below the upper ray of $W_p$, and thus all such points can be handled as in case $(c)$.
Symmetrically, if the projection of $p$ onto $l_i$ is lower than that of $q$, then $q$ lies above the lower ray of
$W_p$, and all such points can be handled as in case $(d)$ (see Figure~\ref{fig:rotateF}).

In other words, case (f) can be handled, with some extra care, as a combination
of cases (c) and (d).


Putting everything together, we obtain:
\begin{theorem} \label{thm:bfs2}
BFS on $VG(P,r)$, in the undirected setup, can be performed in $O(n\log^{O(1)}n)$ randomized expected time.    
\end{theorem}
\subsection{The corresponding bottleneck path problem}

As before, we seek the minimum value $r^*$ for which there exists a path from $s$ to $t$ in $VG(P,r^*)$ 
(or a path of length at most $k$ in the bounded-hop version). This optimization problem can be solved 
using the shrink-and-bifurcate technique of \cite{BFKKS,CH,KKSS}, as above,
using the BFS algorithm of Theorem~\ref{thm:bfs2} as the decision procedure. Here too, the critical values of $r$
are of two kinds: (i) pairwise distances between the points of $P$, and (ii) values at which some point 
$m^+(q)$ or $m^-(p)$ (or their symmetric counterparts) becomes co-vertical or co-horizontal with some 
point of $P$. Both cases are handled as before, first handling critical values of the second kind
using standard parametric search, and then handling critical values of the first kind by the 
shrink-and-bifurcate technique. Omitting the obvious and easy adjustments, we obtain:

\begin{theorem}
The bottleneck path problem, as well as its bounded-hop (RSP) version, for undirected visibility graphs 
involving $n$ directional antennas in the plane, each with an opening angle at least $\alpha$, for 
some constant $\alpha$, can be solved in $O^*(n^{8/7})$ randomized expected time (where the $O^*(\cdot)$
factor also depends on $\alpha$).
\end{theorem}

\section{The case of terrains}
\label{sec:ter}

\subsection{BFS in Visibility graphs over a terrain} \label{sec:ter-bfs}

\begin{figure}[ht]
\centering
\includegraphics[scale=0.8]{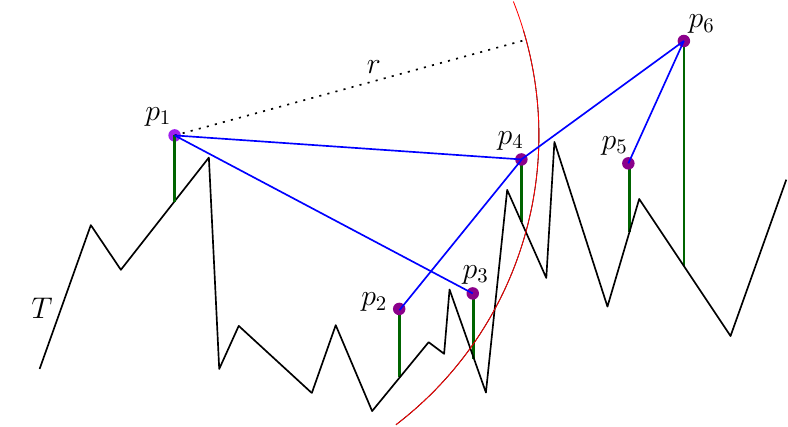}
\caption{
\sf The graph $VG(P,T,r)$. The points $p_1,p_6$ are mutually visible but $|p_1p_6| > r$, while $|p_1p_2| < r$ but $p_1,p_2$ are not mutually visible.} 
\label{fig:terrain}
\end{figure}

Let $T$ be a $1.5$-dimensional terrain, namely an $x$-monotone polygonal path, with $n$ vertices, let $P$ be a set
of $m$ points on or above $T$, and let $r>0$ be a real parameter. Consider the bounded-range visibility graph 
$VG(P,T,r)$, whose vertices are the points of $P$, and, for any $p,q\in P$, there is an edge between $p$ and $q$ 
if and only if they are mutually visible over $T$, meaning that the segment $pq$ lies fully above $T$, and $|pq|$ is at most $r$ (see Figure~\ref{fig:terrain}). 
The decision problem is to determine, for a given $r$, whether $VG(P,T,r)$ contains a path between two designated
points $s,t\in P$, or, more generally, to perform BFS on $VG(P,T,r)$ from a given start point $s$.
The optimization problem, studied in the next subsection, is the bottleneck path problem: Find the smallest value 
$r^*$ for which $VG(P,T,r^*)$ contains a path from $s$ to $t$.
In the bounded-hop version, we are also given an integer 
parameter $k\le m$, and wish, as before, to determine whether
there exists an $s$-$t$ path in $VG(P,T,r)$ with at most $k$ edges, or to find the minimum $r^*$ for which such a path exists.

A simpler version of this problem, without the bounded range restriction, has recently been studied in \cite{KSS}.
Notice that if we only wish to enforce the distance condition, we can use a dynamic Voronoi diagram of $P$, 
as before, at the cost of $O(\log^4m)$ time per search and update. Similarly, if we only wish to enforce unconstrained
visibility (namely, report points $p$ such that $qp$ lies above $T$), we can follow the query procedure given in 
\cite{KSS} (see below for a brief review). The challenge, as before, is to enforce both constraints simultaneously. We describe below a multi-level 
data structure that achieves this goal at a polylogarithmic cost per query, for a total of near-linear running time.  

Our algorithm builds upon the machinery of \cite{KSS}, with nontrivial enhancements that handle the bounded-range
constraint. For the sake of completeness we begin by providing a brief sketch of the technique of \cite{KSS}.

\subparagraph*{A brief review of the technique of \cite{KSS}.}
Let $VG(P,T)$ denote the visibility graph of $P$ over $T$, without any range constraint. We run BFS on $VG(P,T)$,
starting from a point $s$. We maintain the subset $W$ of all the points that the BFS has not reached yet. At each BFS layer
$L_i$, we search with each $q\in L_i$ in $W$ and report the points that are visible from $q$, adding them
to the next layer $L_{i+1}$ and deleting them from $W$.

We use a divide-and-conquer mechanism that repeatedly splits $T$ into a left portion $T_L$ and a right portion $T_R$
by some vertical divider $H$, which is an upward-directed vertical ray emanating from some point on $T$. The set $P$ is split into
the subset $P_L$ of points above $T_L$ and the subset $P_R$ of points above $T_R$. The main step of the algorithm
is to report visibilities across $H$, between points of $P_L$ and points of $P_R$. We prepare a data structure for 
each part separately, and query it with points from the other side of $H$.

Consider the step of querying with points of $P_L$ into $P_R$. We construct, for each point in $P_L\cup P_R$,
the lowest ray that emanates from it and sees $H$. For a point $q\in P_L$ this is the most clockwise such ray, 
and we denote it as $\rho^+(q)$, and for a point $p\in P_R$ this is the most counterclockwise such ray, and we 
denote it as $\rho^-(p)$. (Constructing these rays can be done in polylogarithmic cost per ray; see \cite{KSS}.) 
A point $q\in P_L$ sees a point $p\in P_R$ if and only if the segment $qp$ lies above both rays
$\rho^+(q)$ and $\rho^-(p)$, i.e., $p$ lies above $\rho^+(q)$ and $q$ lies above $\rho^-(p)$. This calls, na\"ively,
for two data structure levels for halfplane range reporting, which would result in an expensive query cost, so
we reduce this task to a single level of halfplane range reporting, using the following machinery.

\begin{figure}[ht]
\centering
\includegraphics[scale=0.6]{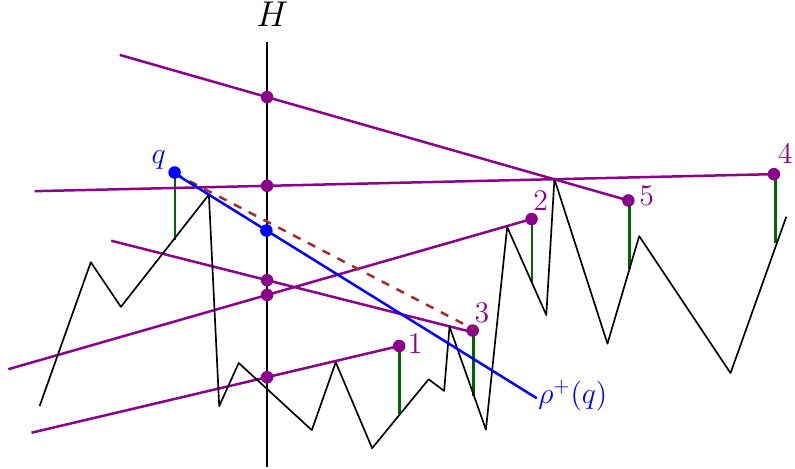}
\caption{Critical rays of $P_R$ with their order by their intercepts on $H$. A point $q\in P_L$ and its critical ray $\rho^+(q)$ are depicted. $q$ sees point 3 because the segment $q3$ lies above both critical rays.} 
\label{fig:critRayOrd}
\end{figure}

We form the intercepts $\rho^-(p)\cap H$ of the rays $\rho^-(p)$, for $p\in P_R$, and sort them, in their $y$-order,
into a one-dimensional range tree $Q$. Given a query point $q\in P_L$, we form the intercept $q_H = \rho^+(q)\cap H$, 
and search with it in $Q$ to obtain a representation of the set of intercepts $Q^+_q$ (resp., $Q^-_q$) that lie
above (resp., below) $q_H$ as the union of $O(\log m)$ canonical subsets. We denote the corresponding sets of $P_R$
as $P_R^+$ and $P_R^-$. See Figure~\ref{fig:critRayOrd}. Clearly, for $p\in P_R^+$ it suffices to test whether $q$ lies above $\rho^-(p)$, and for 
$p\in P_R^-$ it suffices to test whether $p$ lies above $\rho^+(q)$. Thus, with a single (primal or dual) dynamic 
halfplane range reporting, we can report all neighbors of $q$, and delete each neighbor as soon as it is detected.
This results in an overall procedure that runs in $O((m+n)\polylog(m+n))$ time.

This completes our brief overview. See \cite{KSS} for full details, including many steps that we have omitted or glossed over.

\subparagraph*{Our algorithm.}
Going back to our bounded-visibility version of the problem, we seek to report points visible from $q$ 
(across the divider) and at distance at most $r$ from $q$. The main idea is to use the data structure and the query
procedure of \cite{KSS} to obtain the $O(\log m)$ canonical subsets that comprise $P_R^-$ and $P_R^+$. 
Then, using similar ideas to those in the preceding sections, we further partition each set into smaller canonical subsets, so as to accommodate the bounded-range constraint.  
For this we need to extend the data structure of \cite{KSS}, by essentially adding several levels between the first and second levels of the structure. 

\subparagraph*{Reporting the bounded-range visible points from a query $q$ (across the divider).}
This is the main step in the divide-and-conquer procedure. Given a point $q$ on or above $T_L$, say, we first follow the 
query procedure of \cite{KSS}, to obtain the subsets $P_R^-$ and $P_R^+$ of $P_R$, whose intercepts lie below $q_H$
and above $q_H$, respectively, each as the union of $O(\log m)$ canonical subsets, stored at 
$O(\log m)$ respective nodes of the top-level tree of the structure (that handles intercepts on $H$). 

\begin{figure}[ht]
\centering
\includegraphics[scale=0.8]{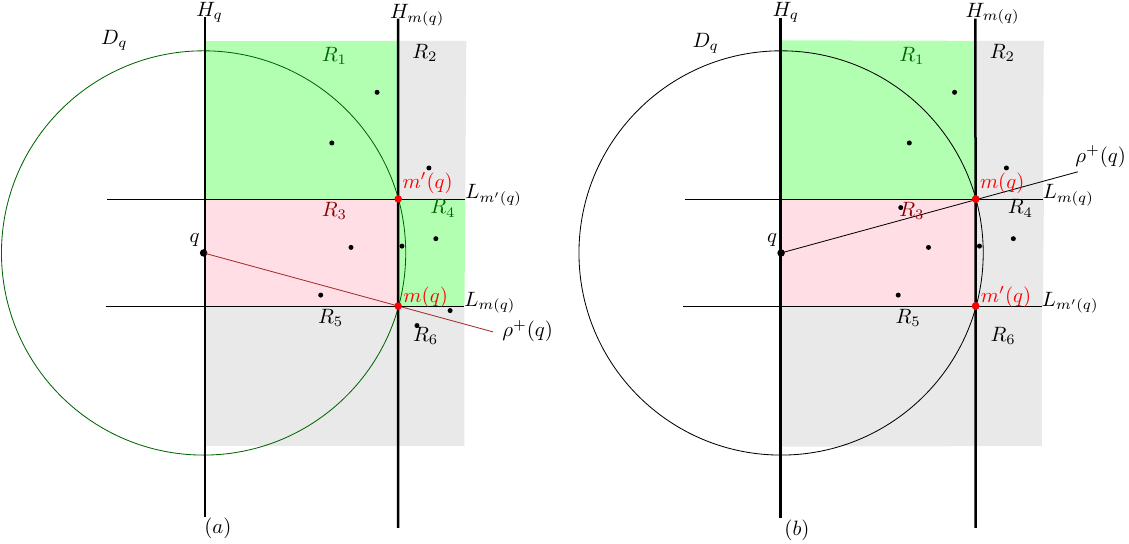}
\caption{
\sf The circle $D_q$, the points $m(q)$, $m'(q)$, and the rectangular partition that they induce. (a) The case where $m'(q)$ is above $m(q)$, so $R_4$ is fully above $\rho^+(q)$. (b) The case where $m(q)$ is above $m'(q)$, so $R_4$ is fully below $\rho^+(q)$.} 
\label{fig:bndRng}
\end{figure}

\subparagraph*{Handling $P_R^-$.}
Consider a canonical subset $P^\eta_R$ of $P_R^-$, and recall that we wish to report all points in $P^\eta_R$ that lie above $\rho^+(q)$, and at distance at most $r$ from $q$. We do this by dividing $P^\eta_R$ into smaller subsets, such that, for each subset we can report all the relevant points by verifying only one of the two (halfplane or distance) conditions. The following procedure actually ignores the divider $H$; it only uses the fact that all the points in $P^\eta_R$ lie to the right of $q$.
Consider the following partition of the halfplane to the right of $q$ into axis-aligned 
rectangles (both bounded and unbounded), and the corresponding
partition of $P^\eta_R$ into subsets $P^\psi_R$, so that each of the subsets is contained in a single rectangle. 
Denote by $m(q)$ the point on $\rho^+(q)$ at distance $r$ from $q$, i.e., $m(q)$ 
is the intersection of $\rho^+(q)$ and the circle $D_q$ of radius $r$ centered at $q$. Let $H_{m(q)}$ be the vertical 
line through $m(q)$, and notice that $H_{m(q)}$ intersects $D_q$ at another point, which we denote as 
$m'(q)$ (unless $\rho^+(q)$ is horizontal). Let $L_{m(q)}$ and $L_{m'(q)}$ be the horizontal lines through $m(q)$ and $m'(q)$, respectively. The lines $H_{m(q)}, L_{m(q)}, L_{m'(q)}$ induce a partition of the halfplane to the right of $q$ into six rectangles $R_1,\ldots, R_6$, as depicted in Figure~\ref{fig:bndRng}. 
We use a two-level orthogonal range tree $P^\eta_R$ to obtain the partition of $P^\eta_R$ into those subsets $P^\psi_R$.
These rectangles, and the respective sets $P_R^\psi$ that they contain, are handled as follows.

$R_1$ is to the left of $H_{m(q)}$ and above both $L_{m(q)}$ and $L_{m'(q)}$. Therefore, all the points in $P_R^{(1)}=P^-_R\cap R_1$ lie above $\rho^+(q)$, so it suffices to report those points at distance at most $r$ from $q$. To do so, we query with $q$ in the corresponding Voronoi diagrams of the $O(\log^2m)$ canonical subsets that comprise $P_R^{(1)}$. Each reported point $p$ at distance at most $r$ from $q$ is added to $L_{i+1}$, and is promptly deleted from all 
the $O(\log^2m)$ structures (Voronoi diagrams and dynamic halfplane range reporting structures) that it participates in.
We stop when no neighbor at that distance is found.

$R_3$ is to the left of $H_{m(q)}$ and between $L_{m(q)}$ and $L_{m'(q)}$, so it is contained in $D_q$. Therefore, all the points in $P_R^{(3)}=P^-_R\cap R_3$ are at distance at most $r$ from $q$, so it suffices to report those points that lie above $\rho^+(q)$. We follow the same approach as before, but now we query the halfplane range reporting structures corresponding to the relevant canonical subsets.

Note that no point below both $L_{m(q)}$ and $L_{m'(q)}$ is visible from $q$ --- it is either too far from $q$ or it lies below $\rho^+(q)$. Therefore, since $R_5 \cup R_6$ is below both $L_{m(q)}$ and $L_{m'(q)}$, we ignore all the points in $P^-_R\cap (R_5\cup R_6)$. Similarly, no point above both $L_{m(q)}$ and $L_{m'(q)}$ and to the right of $H_{m(q)}$ is sufficiently close to $q$. Therefore, we ignore all the points in $P^-_R\cap R_2$. See Figure~\ref{fig:bndRng} for an illustration. 

Finally, $R_4$ is to the right of $H_{m(q)}$ and between $L_{m(q)}$ and $L_{m'(q)}$. We distinguish between two cases, depending on which of $m(q)$ and $m'(q)$ is higher.
If $m'(q)$ lies above $m(q)$ (see Figure~\ref{fig:bndRng}(a)), then, as in the case of $R_1$, all the points in
$P_R^{(4)}=P^-_R\cap R_4$ lie above $\rho^+(q)$, so it suffices to report those points at distance  at most 
$r$ from $q$, and we do it as described for $P_R^{(1)}$. If $m(q)$ lies above $m'(q)$ (see Figure~\ref{fig:bndRng}(b)), 
then all the points in $R_4$ lie below $\rho^+(q)$, therefore we ignore all the points in $P^-_R \cap R_4$. 

The collection of all the reported points, over all canonical subsets, constitutes the first part of the 
output for $q$, namely, all the points of $P^-_R$ that are both visible from $q$ and within distance $r$ from $q$;
all these points are added to the next BFS layer $L_{i+1}$, and promptly deleted from the structures that contain them.

\subparagraph*{Handling $P_R^+$.}
Consider now a canonical subset $P^\eta_R$ of $P_R^+$. Here we wish to report all points $p$ within distance at most
$r$ from $q$, such that $q$ lies above $\rho^-(p)$. Again, 
we do this by dividing $P^\eta_R$ into smaller subsets, such that, for each subset we can report all the relevant points by verifying only one of the two (halfplane or distance) conditions.

\begin{figure}[ht]
\centering
\includegraphics[scale=0.8]{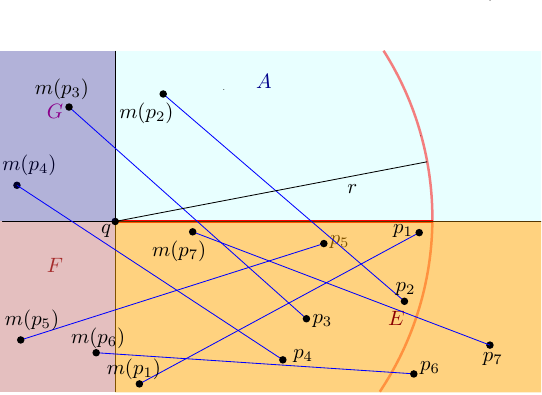}
\caption{
\sf The partition of the plane defined by $q$. Points $p\in E$ and the corresponding segments $pm(p)$.}
\label{fig:bndRng2}
\end{figure}

Consider first the case of points in $P^\eta_R$ that lie below $q$. We describe a solution for this case, which is similar in flavor to the solution for the second case (i.e., $\alpha_p\le\alpha_q$) in Section~\ref{subsec:antennas-bfs-undir}. 
Consider the partition of the plane into the four quadrants defined by $q$, which we denote $A$, $E$, $F$ and $G$, as depicted in Figure~\ref{fig:bndRng2}. Notice that here we consider points that lie below $q$, i.e. in $E$.

Let $m(p)$ be the point that lies at distance $r$ from $p$ along $\rho^-(p)$, and
distinguish between several subcases, depending on which of the four quadrants defined by $q$ contains $m(p)$. 

If $m(p)$ lies in the southwest quadrant $F$ (see points $p_5$ and $p_6$ in Figure~\ref{fig:bndRng2}), then clearly $q$ lies above $\rho^-(p)$, and it suffices
to test whether $|pq|\le r$, by searching in the suitable Voronoi diagrams.

If $m(p)$ lies in the northeast quadrant $A$ (see point $p_2$ in Figure~\ref{fig:bndRng2}), then clearly $q$ lies below $\rho^-(p)$, so this case can be ignored.  

Consider the case where $m(p)$ lies in the northwest quadrant $G$ (see points $p_3$ and $p_4$ in Figure~\ref{fig:bndRng2}). Then both segments $m(p)q$ and $qp$ have
negative slopes, implying that the angle $\angle m(p)qp$ between them is obtuse. Therefore, $|pq| < |pm(p)| = r$,
so it suffices to test whether $q$ lies above $\rho^-(p)$.
(This step is implemented in the 
dual plane, as a halfplane reporting query with the line dual to $q$ amid the points dual to the lines containing $\rho^-(p)$ for points $p$ in the appropriate set of yet undiscovered points.)

Finally, consider the case where $m(p)$ lies in the southeast quadrant, i.e., in $E$ (see points $p_1$ and 
$p_7$ in Figure~\ref{fig:bndRng2}). Here we claim that if $|pq| \le r$, then $q$ lies above $\rho^-(p)$. Thus, it 
suffices to search in the suitable Voronoi diagrams. Indeed, if $q$ lies below $\rho^-(p)$, then
$pm(p)$ necessarily points upwards (since $m(p) \in E$) and so does $m(p)q$. Therefore, the angle $\angle qm(p)p$ is obtuse, implying that $|pq| > |pm(p)| = r$. 

To recap, knowing the quadrant containing $m(p)$ allows us either to discard $p$ altogether as a possible neighbor 
of $q$, or to determine which of the two reporting tasks (halfplane range reporting in the dual plane or searching in Voronoi diagrams) 
suffices to determine whether $p$ is a neighbor.

Consider now the case of the points in $P^\eta_R$ that lie above $q$. The solution for this case is similar to the 
solution for the previous case. That is, we continue to consider the partition of the plane into quadrants $A$, 
$E$, $F$ and $G$, except that now we consider points that lie above $q$, i.e., in $A$. A similar analysis, which is
based on the location of the point $m(p)$, shows that in each of the four subcases we can resort to only one of 
the two (halfplane or distance) conditions, or discard $p$ completely. Specifically, if $m(p)$ lies in $G$, then we 
can ignore $p$, since in this case, $q$ cannot lie above $\rho^-(p)$. If $m(p)$ lies in $A$, then we can also ignore 
$p$, since if the ray $\rho^-(p)$ points upwards, then clearly $q$ cannot lie above it, and if it points downwards, 
then $p$ is too far from $q$ (because, similar to a previous argument, the angle $\angle qm(p)p$ is obtuse). If $m(p)$ lies in $E$, then it suffices to test 
whether $|pq| \le r$, since in this case $q$ is surely above $\rho^-(p)$. Finally, if $m(p)$ lies in $F$, then surely 
$|pq| \le r$, and it suffices to test whether $q$ lies above $\rho^-(p)$.

The collection of all the reported points, over all canonical subsets, constitutes the second part of the output for $q$, namely, all the points of $P^+_R$ that are both visible from $q$ and within distance $r$ from $q$.

The cost of a query is dominated by the cost of searching in, and updating, the various dynamic Voronoi diagrams.
Since the number of canonical sets that a query processes is $O(\log^5m)$ (because they are collected over five 
levels of the structure), the cost of a query is $O(\log^9 m)$. Hence the overall cost of the BFS is $O(m \log^9 m \log n)$, plus the preparatory step, as in \cite{KSS}, which takes $O(n\log n + m\log^2 n)$ time.

In conclusion, we have shown:
\setcounter{theorem}{4}

\begin{theorem} \label{thm:bfs-ter}
BFS on $VG(P,T,r)$, for a $1.5$-dimensional terrain $T$ with $n$ vertices and a set $P$ of $m$ points over $T$, can be performed in $O(n\log n + m\log^2 n + m\log^9 m \log n)$ randomized expected time.   
\end{theorem}

\subsection{The corresponding bottleneck path problem}
\label{sec:ter-bn}
We seek the minimum value $r^*$ for which there exists a path from $s$ to $t$ in $VG(P,T,r^*)$. As in the preceding 
sections, this optimization problem can be solved using the shrink-and-bifurcate technique of \cite{BFKKS,KKSS},
using the BFS algorithm of Theorem~\ref{thm:bfs-ter} as the decision procedure. Here too, the critical values of $r$
are of two kinds: (i) pairwise distances between the points of $P$, and (ii) values at which some point $m(q)$ 
or $m(p)$ becomes co-vertical or co-horizontal with some point of $P$. We handle each kind separately, as before. 
We begin the procedure by sorting all the points of $P$ into the two-dimensional orthogonal range tree $Q$. 
We then locate the $y$-coordinates of the points $m(q)$ and $m(p)$ in the primary tree, to obtain the
canonical subsets that represent the points to the left and to the right of each $m(q)$ and $m(p)$.

As before, simulation of this step, with the unknown $r^*$, can be performed using standard parametric search.
Similarly, for each canonical set $T_v$ that we obtain for some $m(q)$ or $m(p)$, we search with the $x$-coordinate 
of $m(q)$ or $m(p)$ in $T_v$. This step too can be done in parallel, over all points $q$, all nodes $v$, 
and the $O(\log m)$ steps of each search.

Hence, when this part of the simulation ends, all critical values of type (ii) are out of the interval that contains
$r^*$, and each comparison that involves them can be resolved, independently of $r^*$.

For critical values of type (i), we use the shrink-and-bifurcate technique of \cite{BFKKS,KKSS} mentioned above.
Using the \cite{CH} implementation for the  shrinking part of the procedure implies that the overall technique yields a procedure that runs in $O^*(m^{8/7})$ randomized expected time. That is, we have:
\begin{theorem}
The bottleneck path problem for bounded-range visibility graphs over a terrain, as well as its bounded-hop (RSP) version,
can be solved in $O^*(n + m^{8/7})$ randomized expected time. 
\end{theorem}
\bibliographystyle{plain}
\bibliography{refs}

\end{document}